\documentstyle[prl,aps,amsfonts,multicol]{revtex}
\newcommand{\trace}{\mathop{\rm Tr}\nolimits}
\newcommand{\qed}{\hfill$\Box$}
\newtheorem{theorem}{Theorem}

\draft
\title{Local filtering operations on two qubits}
\author{Frank Verstraete, Jeroen Dehaene, Bart De Moor\cite{FJBmail}}
\address{Katholieke Universiteit Leuven,
Department of Electrical Engineering, Research Group SISTA\\
Kard. Mercierlaan 94, B-3001 Leuven, Belgium }
\date{\today}
\begin{document}

\pagestyle{plain} \pagenumbering{arabic}

\maketitle
\begin{abstract}
We consider one single copy of a mixed state of two qubits and
investigate how its entanglement changes under local quantum
operations and classical communications (LQCC) of the type
$\rho'\sim (A\otimes B)\rho(A\otimes B)^{\dagger}$. We consider a
real matrix parameterization of the set of density matrices and
show that these LQCC operations correspond to left and right
multiplication by a Lorentz matrix, followed by normalization. A
constructive way of bringing this matrix into a normal form is
derived. This allows us to calculate explicitly the optimal local
filterin operations for concentrating entanglement. Furthermore we
give a complete characterization of the mixed states that can be
purified arbitrary close to a Bell state. Finally we obtain a new
way of calculating the entanglement of formation.
\end{abstract}

\begin{multicols}{2}[]
\narrowtext Entanglement of two separated quantum systems implies
that there are non-local correlations between them. This feature
of non-locality has found practical applications in quantum
information theory (see for example \cite{Lo}). Most applications
require that both parties share maximally entangled states. A
realistic preparation and transmission of entangled states
however yields mixed states. Therefore Bennett et al.
\cite{Bennett} proposed a protocol which allows to obtain
asymptotically a nonzero number of maximally entangled pure
states by carrying out collective measurements on a large number
of copies of entangled mixed states. Their scheme however
required that the fidelity of the mixed states exceeded 1/2. The
Horodecki's subsequently showed how mixed states of arbitrary
fidelity could be purified by applying first a filtering
operation on each copy separately \cite{Horodecki2}. Linden et al.
\cite{Linden} then asked the question if it is possible to obtain
singlets out of mixed states by allowing only local operations on
each copy separately. While this is possible for pure states, they
proved that this is impossible in general for mixed states
\cite{Linden,Kent1}, as the best state one can obtain is a Bell
diagonal state \cite{Kent2}. The Horodecki's however gave an
example of a mixed state that could be purified arbitrary close
to a singlet state by a process called quasi-distillation
\cite{Horodecki3}.

We shed new light on those results by observing that filtering
operations on two qubits correspond to Lorentz transformations on
a real parameterization of their density matrix. Using Lorentz
transformations, this real parameterization can be brought into
one of two types of normal forms, thus giving a characterization
of all states that can be transformed into each other by local
operations. Our scheme also yields a new way of calculating the
entanglement of formation \cite{Wootters}, with as a by-product a
simple proof of the necessity and sufficiency of the partial
transpose criterion of Peres \cite{Peres,Horodecki}. The main
result of this letter however is the fact that we provide a
constructive way of finding the optimal POVM's for concentrating
the entanglement. We show that there exist two classes of states
corresponding to the two normal forms : the ones that can be
brought into Bell diagonal form leaving the rank of the density
matrix constant, and the ones that can asymptotically be brought
into Bell diagonal form with lower rank. This last class contains
a subclass of mixed states that can be purified arbitrary close
to the singlet state.

In this letter we will consider the filtering operations
\begin{equation} \rho'=\frac{(A\otimes B)\rho(A\otimes
B)^{\dagger}}{\trace((A\otimes B)\rho(A\otimes
B)^{\dagger})}\label{LQCC}\end{equation} where $A^{\dagger}A\leq
I_2$, $B^{\dagger}B\leq I_2$. As a local projective measurement
destroys all entanglement, we will only consider the cases ${\rm
det}(A)\neq 0$ and ${\rm det}(B)\neq 0$. Let us now calculate how
the entanglement of formation (EoF) changes under these local
operations. The EoF of a two qubit system can be calculated as a
convex monotonously increasing function of the concurrence
\cite{Wootters}. As shown in \cite{Verstraete1}, the concurrence
of $\rho$ is given by $\max(0,\tau_1-\tau_2-\tau_3-\tau_4)$ with
$\{\tau_i\}$ the singular values of
$X^T(\sigma_y\otimes\sigma_y)X$ with $\rho=XX^\dagger$. Under the
filtering operations we have $X'=(A\otimes
B)X/\sqrt{\trace(A^\dagger A\otimes B^\dagger B\rho)}$. As
$(A\otimes B)^T(\sigma_y\otimes\sigma_y)(A\otimes
B)=\det(A)\det(B)(\sigma_y\otimes\sigma_y)$, this proves the
following theorem:
\begin{theorem}
Under the filtering operations (\ref{LQCC}), the concurrence
changes as
\begin{equation}C'=C\frac{|\det(A)||\det(B)|}{\trace(A^\dagger
A\otimes B^\dagger B\rho)}.\label{C}\end{equation}
\label{t3}\end{theorem}

It will turn out very usefull to introduce the real and linear
parameterization of the density matrix \cite{Mahler}
\begin{equation}\rho=\frac{1}{4}\sum_{i,j}R_{ij}\sigma_i\otimes\sigma_j\end{equation}
where the summation extends from 0 to 3 and with $\sigma_0$ the
2x2 identity matrix and $\sigma_1,\sigma_2,\sigma_3$ the Pauli
spin matrices. Below we will often leave out the normalization of
$\rho$ and $R$. Note that normalization of $R$ is very simple
since $R_{0,0}=\trace(\rho)$.

We will now prove how $R$ transforms under the LQCC operations
(\ref{LQCC}).
\begin{theorem}
The {\rm 4x4} matrix $R$ with elements
$R_{ij}=\frac{1}{2}\trace(\rho(\sigma_i\otimes\sigma_j))$
transforms, up to normalization, under LQCC operations
(\ref{LQCC}) as \begin{equation}R'=L_ARL_B^T\end{equation} where
$L_A$ and $L_B$ are proper orthochronous Lorentz transformations
given
by \begin{eqnarray}&L_A&=T(A\otimes A^*)T^{\dagger}/|{\rm det}(A)|\\
&L_B&=T(B\otimes B^*)T^{\dagger}/|{\rm det}(B)|\\
T&=&\frac{1}{\sqrt{2}}\left(\begin{array}{cccc}
1&0&0&1\\0&1&1&0\\0&i&-i&0\\1&0&0&-1\end{array}\right).\end{eqnarray}
\label{t1}\end{theorem} This theorem can be proven by introducing
the matrix $\tilde{\rho}_{kl,k'l'}=\rho_{kk',ll'}$ and noting that
$R=4T\tilde{\rho}T^{T}$. It is easy to check that under the LQCC
operations (\ref{LQCC}) $\tilde{\rho}$ transforms as
$\tilde{\rho}'=(A\otimes A^*)\tilde{\rho}(B\otimes B^*)^T$.
Therefore $R$ transforms as $R'=L_A R L_B^T|{\rm det}(A)||{\rm
det}(B)|$ with $L_A=T(A\otimes A^*)T^\dagger/|{\rm det}(A)|$,
$L_B=T(B\otimes B^*)T^\dagger/|{\rm det}(B)|$.  Using the
identities  $A\sigma_y A^T={\rm det}(A)\sigma_y$ and $T^\dagger M
T^*=-\sigma_y\otimes\sigma_y$ with $M$ the matrix associated with
the Lorentz metric $ M={\rm diag}[1,-1,-1,-1]$, it is easily
checked that $L_A M L_A^T=M=L_B ML_B^T$. Furthermore the
determinant of $L_A$ and $L_B$ is equal to +1, and the $(0,0)$
element of $L$ is positive, which completes the proof.\qed

As the complex 2$\times$2 matrices with determinant one indeed
form the spinor representation of the Lorentz group,  there is a
1 to 2 correspondence between each $L_A$ and $A/\sqrt{{\rm
det}(A)}$. It is interesting to note that when both $A$ and $B$
are unitary, the theorem reduces to the well-known fact
\cite{Mahler} that the rows and columns of $R$ transform under
$SO(3)$, which is indeed a subgroup of the Lorentz group.

With the above theorem in mind, a natural question is to find a
decomposition of $R$ as $R=L_1\Sigma L_2^T$ with $\Sigma$
diagonal and $L_1,L_2$ proper orthochronous Lorentz
transformations. This would be the analogue of a singular value
decomposition but now in the Lorentz instead of the Euclidean
metric.
\begin{theorem}
The {\rm 4x4} matrix $R$ with elements
$R_{ij}=\trace(\rho\sigma_i\otimes\sigma_j)$ can be decomposed as
\begin{equation}R=L_1\Sigma L_2^T\end{equation} with $L_1,L_2$ proper orthochronous
Lorentz transformations, and $\Sigma$ either of diagonal form
$\Sigma={\rm diag}[s_0,s_1,s_2,s_3]$ with $s_0\geq s_1\geq s_2\geq
|s_3|$, either of the form
\begin{equation}\Sigma=\left(\begin{array}{cccc}a&0&0&b\\0&d&0&0\\0&0&-d&0\\c&0&0&a+c-b\end{array}\right)\label{jb}\end{equation}
with $a,b,c,d$ real. \label{t2}\end{theorem} The proof of this
theorem is quite technical. It heavily depends on results on
matrix decompositions in spaces with indefinite metric
\cite{Gohberg}. We first introduce the matrix $C=MRMR^T$ which is
$M$-selfadjoint. Using theorem (5.3) in \cite{Gohberg}, it
follows that there exist matrices $X$ and $J$ with $C=X^{-1}JX$,
$J$ consisting of a direct sum of real Jordan blocks and
$XMX^T=N_J$ with $N_J$ a direct sum of symmetric nxn matrices of
the form $[S_{ij}]=\pm[\delta_{i+j,n+1}]$ with $n$ the size of the
corresponding Jordan block. Using Sylvester's law of inertia,
there exists orthogonal $O_J$ such that $N_J=O_J^TMO_J$. It is
then easy to check that $O_JX=L_1^T$ is a Lorentz transformation.
Therefore the relations $C=MRMR^T=ML_1MO_JJO_J^TL_1^T$ hold.
Multiplying left by $M$, Sylvester's law of inertia implies that
there exist a matrix $\Sigma$ with the same rank as $J$ such that
$MO_JJO_J^T=\Sigma M\Sigma^T$. Therefore we have the relation
$RMR^T=L_1\Sigma M\Sigma^TL_1^T$. If $R$ has the same rank as
$RMR^T$, this relation implies that there exists a Lorentz
transformation $L_2$ such that $R=L_1\Sigma L_2^T$.

Let us now investigate the possible forms of $\Sigma$. As
$N_J=O_J^TMO_J$ has signature $(+---)$, $J$ can only be a direct
sum of the following form: 4 1x1 blocks; 1 orthogonal 2x2 block
and 2 1x1 blocks; 1 2x2 Jordan block and 2 1x1 blocks; 1 3x3
Jordan block and 1 1x1 block. Noting the eigenvalues of $C$ as
$\{\lambda_i\}$, it is easy to verify that a "square root"
$\Sigma$ in the four cases is respectively given by
\begin{enumerate}
\item $\Sigma={\rm
diag}[\sqrt{|\lambda_0|},\sqrt{|\lambda_1|},\sqrt{|\lambda_2|},\sqrt{|\lambda_3|}]P$
with $P$ a permutation matrix permutating the first column with
one other column;
\item $\Sigma={\rm
diag}\left[\sqrt{|\lambda_0|}\left(\begin{array}{cc}\cos(\phi)&\sin(\phi)\\
\sin(\phi)&-\cos(\phi)\end{array}\right),\sqrt{|\lambda_2|},\sqrt{|\lambda_3|}\right]$;
\item $\Sigma={\rm
diag}\left[\left(\begin{array}{cc}a&b\\c&a+c-b\end{array}\right),\sqrt{|\lambda_2|},\sqrt{|\lambda_3|}\right]$;
\item $\Sigma={\rm diag}\left[\left(\begin{array}{ccc}a
&0&0\\b&\sqrt{a^2+b^2}&0\\0&\frac{-ab}{\sqrt{a^2+b^2}}&\frac{a^2}{\sqrt{a^2+b^2}}\end{array}\right),\sqrt{|\lambda_3|}\right]$
with $a=\sqrt{|\lambda_0|}$ and $b=-1/\sqrt{2|\lambda_0|}$.
\end{enumerate}
Now we go back to the relation $R=L_1^T\Sigma L_2$. $L_1$ and
$L_2$ can be made proper and orthochronous by absorbing factors
$-1$ into the rows and colums of $\Sigma$ yielding $\Sigma'$.
Theorem (2) now implies that this $\Sigma'$ corresponds to an
unnormalized physical state, which means that $\rho'$
corresponding to $\Sigma'$ has no negative eigenvalues. It is
easy to show that this requirement excludes cases 2 and 4 of the
possible forms of $\Sigma$. The third case corresponds to
(\ref{jb}). Furthermore in the first case the permutation matrix
has to be the identity and
$|\lambda_0|\geq\max(|\lambda_1|,|\lambda_2|,|\lambda_3|)$.
Multiplying left and right by proper orthochronous Lorentz
transformations, the elements $\{s_i\}$ of this diagonal $\Sigma$
can always be ordered as $s_0\geq s_1\geq s_2\geq |s_3|$.

The case where the rank of $C$ is lower then the rank of $R$
still has to be considered. This is only possible if the rowspace
of $R$ has an isotropic subspace $Q$ for which $QMQ^T=0$. Some
straightforward calculations reveal that the only physical states
for which this hold have normal form (\ref{jb}) with $a=b=c$ and
$d=0$ or $a=b$ and $c=d=0$. This completes the proof.\qed

The two normal forms can be computed very efficiently by
calculating the Jordan canonical decomposition of $C=MRMR^T$ and
of $C'=MR^TMR$. It is easy indeed to show that for example in the
case of diagonalizable $R$ the eigenvectors of $C$ form a Lorentz
matrix, and $|s_i|=\sqrt{\lambda_i(C)}$. Note that we always order
the diagonal elements such that $s_0\geq s_1\geq s_2\geq|s_3|$.

States that are diagonal in $R$ correspond to (unnormalized)
Bell-diagonal states with ordered eigenvalues \begin{eqnarray}\lambda_1&=&(s_0+s_1+s_2-s_3)/4\\
\lambda_2&=&(s_0+s_1-s_2+s_3)/4\\
\lambda_3&=&(s_0-s_1+s_2+s_3)/4\\
\lambda_4&=&(s_0-s_1-s_2-s_3)/4,\label{ebd}\end{eqnarray} whereas
states of type (\ref{jb}) correspond to the rank deficient states
\begin{equation}\rho=\frac{1}{2}\left(\begin{array}{cccc}a+c&0&0&d\\0&0&0&0\\0&0&b-c&0\\d&0&0&a-b\end{array}\right)\label{n2}.\end{equation}
For both cases it is easy to calculate the entanglement of
formation analytically, respectively given by \cite{Hill}
$C=\max(0,(\lambda_1-\lambda_2-\lambda_3-\lambda_4)/(\lambda_1+\lambda_2+\lambda_3+\lambda_4))=\max(0,(-s_0+s_1+s_2-s_3)/(2s_1))$
and $C=\max(0,|d|/a)$.

Let us now consider an arbitrary state $\rho$ with corresponding
$R$. Combining theorem (1),(2) and (3), it follows that the
concurrence of $\rho$ is equal to the concurrence of the state
corresponding to $\Sigma$ multiplied by $R_{00}$. We have
therefore proven:
\begin{theorem} Given a state $\rho$ and associated with this
state $R=L_1\Sigma L_2^T$, then the concurrence of $\rho$ is
given by $C=\max(0,(-s_0+s_1+s_2-s_3)/2)$ or by $C=\max(0,|d|)$
depending on the normal form $\Sigma$.\label{t10}
\end{theorem}

We thus have obtained a new method of calculating the entanglement
of formation of a system of two qubits. Interestingly, it turns
out that this characterization relates the concepts of
entanglement of formation and of partial transposition
\cite{Peres}. Let us therefore define
$R^{PT}_{ij}=\trace(\rho^{PT}\sigma_i\otimes\sigma_j)$, which
changes the sign of the third column of $R$. In the case of
diagonal normal form of $R$, it is readily verified that the
normal form of $R^{PT}$ equals that of $R$ except for the last
element where $s_3^{PT}=-s_3$. Retransforming $\Sigma^{PT}$ to the
$\rho^{PT}$-picture, we see that the corresponding Bell-diagonal
partial transposed state has minimal eigenvalue $s_0-s_1-s_2+s_3$.
We readily recognize the expression of the concurrence of theorem
(\ref{t10}) and therefore this eigenvalue is negative if and only
if the concurrence exceeds 0. Moreover we know that $\rho^{PT}$
is related to this Bell-diagonal state by some similarity
transformation $A\otimes B$ which cannot change the signature of
a matrix due to the inertia law of Sylvester. In the case of
normal form (\ref{jb}), an analogue reasoning shows that
$\rho^{PT}$ has a negative eigenvalue if and only if $|d|>0$,
which again is necessary and sufficient to have entanglement.
This completes the proof of:
\begin{theorem} Given a system of two qubits, this state is
separable if and only if its partial transpose has a negative
eigenvalue.\label{t11}\end{theorem} Although this result was
already proven by Horodecki \cite{Horodecki}, we believe the
previous derivation is of interest as it connects the
entanglement measures concurrence and negativity. Using this
formalism, it indeed becomes possible to prove that the
concurrence always exceeds the negativity, and it is furthermore
possible to find a complete characterization of all states with
maximal and minimal negativity for given concurrence
\cite{Verstraete3}. This is important because in the two qubit
case the negativity is a measure of the robustness of
entanglement against noise.

Next we want to solve the problem of finding the POVM such as to
have a non-zero chance to produce a new state with highest
possible entanglement. From equation (\ref{C}), the maximum EoF
is obtained with $A,B$ minimizing the expression
$\trace(A^\dagger A\otimes B^\dagger B\rho)/(|{\rm det(A)}{\rm
det}(B)|$. Absorbing the factors $|{\rm det}(A)|$ and $|{\rm
det}(B)|$ into $A$ and $B$, it is sufficient to consider $A$ and
$B$ with determinant 1. In the $R$-picture, the optimization is
then equal to minimizing the $(0,0)$ element of $R=L_1\Sigma
L_2^T$ by appropriate $L_A,L_B$. Absorbing $L_1$ and $L_2$ into
$L_A'=L_AL_1^TM$ and $L_B'=L_BL_2^TM$, this is equivalent to
finding the optimal vectors $l_A$ and $l_B$ such that
$l_a^T\Sigma l_B$ is minimized under the constraints
$l_A^TMl_A=1=l_B^TMl_B$.

Let us first consider the case of diagonal $\Sigma$ with elements
$s_0\geq s_1\geq s_2\geq |s_3|$. Parameterizing $l_A$ as
$(\sqrt{1+\|\vec{x}\|^2},\vec{x})$ and $l_B$ as
$(\sqrt{1+\|\vec{y}\|^2},\vec{y})$, the following inequalities
hold: $l_A^T\Sigma l_B\geq
s_0\sqrt{1+\|\vec{x}\|^2}\sqrt{1+\|\vec{y}\|^2}-s_1\|\vec{x}\|\|\vec{y}\|\geq
s_0$. Therefore the concurrence will be maximized for
$\vec{x}=\vec{y}=0$, leaving $\Sigma$ into diagonal form.
Collecting the previous results, it follows that if $R$ is
diagonalizable, the state with maximal concurrence that can be
obtained from it by single copy LQCC operations is the one
corresponding to $\Sigma$ which is a Bell diagonal state. This is
in complete accordance with the results of Kent et al.
\cite{Kent2}. The optimal $A$ and $B$ are thus given by the 2x2
matrices corresponding to $L_1^TM$ and $L_2^TM$. The optimal POVM
can then be obtained by dividing $A$ and $B$ by their largest
singular value such that $A^{\dagger}A\leq 1$ and
$B^{\dagger}B\leq 1$, followed by calculating the square roots
$A_c=\sqrt{I_2-A^{\dagger}A}$ and $B_c=\sqrt{I_2-B^{\dagger}B}$
which are rank 1. The optimal POVM's to be performed on the two
qubits are then given by $\{A,A_c\}$ and  $\{B,B_c\}$
respectively. Note that the probability of measuring $(A,B)$ is
given by the inverse of the gain in concurrence divided by the
product of the largest singular values of $A$ and $B$, and that
the rank of the Bell diagonal state is equal to the rank of the
original state.

If $\Sigma$ is of the form (\ref{jb}) however, things get more
complicated. An analogous reasoning as in the diagonal case leads
to the conclusion that $l_A$ and $l_B$ are vectors associated
with the Lorentz transformations bringing (\ref{jb}) into diagonal
form. This is however only possible in the limit where $l_A$ and
$l_B$ contain factors $\lim_{t\rightarrow
\infty}[\sqrt{1+t^2},0,0,t]$ and $\lim_{t\rightarrow
\infty}[\sqrt{1+t^2},0,0,-t]$ respectively. This indeed allows to
bring $R$ asymptotically into diagonal form with diagonal
elements given by $[\sqrt{(a-b)(a+c)},d,-d,\sqrt{(a-b)(a+c)}]$ and
off-diagonal elements of order $1/t^2$, yielding a state
infinitesimally close to a Bell diagonal state. The probability to
get this state during a measurement of the optimal POVM however
scales as $\lim_{t\rightarrow\infty}1/t^2$. This is equivalent to
the quasi-distillation protocol by Horodecki \cite{Horodecki}. In
this limit of $t\rightarrow\infty$ the rank of the new state is
less than the original one, and its concurrence is given by
$|d|/\sqrt{(a-b)(a+c)}$.

In the case where $a-b=a+c=|d|$ we are therefore able to create a
state arbitrary close to the singlet state. Therefore the only
mixed states that can be quasi-purified to the singlet state by
single copy LQCC operations are the rank two states having normal
form (\ref{jb}) with $a-b=a+c=|d|$.

In conclusion, we obtained new insight into the problem of local
filtering on one copy of two qubits by introducing the notion of
Lorentz transformations on a real matrix parameterization of their
density matrix. This matrix can be brought into one of two types
of normal forms. These normal forms contain all the information
about the entanglement of formation and reveal an elementary
connection between concurrence and the partial transpose
criterion of Peres. Moreover, this new formalism enabled us to
derive in a constructive way the optimal local filtering
operations for concentrating entanglement on an arbitrary mixed
state of two qubits. This could be of great interest in
constructing optimal distillation protocols. We showed that
states of the first type can be locally transformed into a Bell
diagonal state of the same rank with finite probability, whereas
states of the second kind can asymptotically be transformed into
Bell diagonal states with lower rank. This last class is of
special interest as is contains the mixed states that can be
transformed arbitrary close to the singlet state.

\vspace{.5 cm} Frank Verstraete is PhD student, Jeroen Dehaene is
postdoctoral researcher and Bart De Moor is full Professor at the
K.U.Leuven. {\small This work is supported by several
institutions:  1. the Flemish Government: a. Research Council
K.U.Leuven : Concerted Research Action Mefisto-666; b. the FWO
projects G.0240.99, G.0256.97, and Research Communities: ICCoS
and ANMMM; c. IWT projects: EUREKA 2063-IMPACT, STWW; 2. the
Belgian State: a. IUAP P4-02 and IUAP P4-2; b. Sustainable
Mobility Programme - Project MD/01/24; 3. the European
Commission: a. TMR Networks: ALAPEDES and System Identification;
b. Brite/Euram Thematic Network : NICONET;3. Industrial Contract
Research : ISMC, Data4S, Electrabel, Laborelec, Verhaert,
Europay; The scientific responsibility is assumed by the authors.}

\end{multicols}

\begin{thebibliography}{99}
\bibitem[*]{FJBmail}
\{frank.verstraete,jeroen.dehaene,bart.demoor\}\\
@esat.kuleuven.ac.be
\end{thebibliography}
\end{document}